\documentclass[conference]{IEEEtran}
\IEEEoverridecommandlockouts
\usepackage[utf8]{inputenc}
\IEEEoverridecommandlockouts

\usepackage{cite}
\usepackage{amsmath,amssymb,amsfonts}
\usepackage{algorithmic}
\usepackage{graphicx}
\usepackage{textcomp}
\def\BibTeX{{\rm B\kern-.05em{\sc i\kern-.025em b}\kern-.08em
    T\kern-.1667em\lower.7ex\hbox{E}\kern-.125emX}}
    
\usepackage{listings}
\usepackage{array} 
\usepackage{ragged2e}
\usepackage{enumitem}
\usepackage{url}
\usepackage{wrapfig}
\usepackage{balance}
\usepackage{multirow}
\usepackage{makecell}
\usepackage{arydshln}
\usepackage{flushend}

\usepackage[utf8]{inputenc}
\usepackage{fourier}

\usepackage[most]{tcolorbox}
\usetikzlibrary{calc}
\usepackage{todonotes}

\newcommand{\fig}[1]{Fig.~\ref{fig:#1}}

\begin{document}

\title{Microservice Architecture Reconstruction and Visualization Techniques: A Review
\thanks{This material is based upon work supported by the National Science Foundation under Grant No. 1854049, grant from Red Hat Research, and Ulla Tuominen (Shapit).}
}

\author{\IEEEauthorblockN{Tomas Cerny}
\IEEEauthorblockA{\textit{\hspace{1em}Department of Computer Science\hspace{1em}} \\
\textit{Baylor University}\\
Waco, Texas, United States \\
tomas\_cerny@baylor.edu}
\and
\IEEEauthorblockN{Amr S. Abdelfattah}
\IEEEauthorblockA{\hspace{1em}\textit{Department of Computer Science\hspace{1em}} \\
\textit{Baylor University}\\
Waco, Texas, United States \\
amr\_elsayed1@baylor.edu}
\and
\IEEEauthorblockN{Vincent Bushong}
\IEEEauthorblockA{\hspace{1em}\textit{Department of Computer Science\hspace{1em}} \\
\textit{Baylor University}\\
Waco, Texas, United States \\
vinbush@gmail.com}
\and
{\,\hspace{3.6em}\,}
\and
\IEEEauthorblockN{Abdullah Al Maruf}
\IEEEauthorblockA{\textit{\hspace{4.6em}Department of Computer Science\hspace{4.6em}} \\
\textit{Baylor University}\\
Waco, Texas, United States \\
maruf\_maruf1@baylor.edu}
\and
\IEEEauthorblockN{Davide Taibi}
\IEEEauthorblockA{\hspace{1.6em}\textit{CloudSEA.AI Group\hspace{1.6em}} \\
\textit{Tampere University}\\
Tampere, FI-33720, Finland \\
davide.taibi@tuni.fi}
\and
{\,\hspace{3.6em}\,}
}





\maketitle
\begin{abstract}

Microservice system solutions are now mainstream. The older microservices-based systems are not more than 15 years old, and their architecture is by far different than the one originally designed because of several changes applied to the systems due to the implementation of new features and bug fixing. 
The evolution of these legacy systems is therefore subjected to degradation. 
One of the most important methods to identify degradation is being able to reconstruct the software architecture of a system based on the current system running in production. 
Different methods have been proposed in the past: methods based on the static analysis of the source code of the microservices and methods based on the analysis of the log traces collected at runtime. 
Both static and dynamic analysis-based methods have their pros ad cons. 
In this work, we review the existing technologies for static and dynamic architectural reconstruction and related tools adopted to visualize the reconstructed architecture. 
The result of this work can be useful both to practitioners and researchers that can further develop these methods to provide better support for architectural degradation.

\end{abstract}


\section{Introduction} \label{sec:introduction}

The software architecture provides the major perspective for the system's development and design. The software architecture serves as the blueprint for systems. The system architecture must often be reconstructed to determine if the system was built accurately. Various works described such a process. O'Brien's report \cite{techRep} defines architecture reconstruction as "the process by which the architecture of an implemented system is obtained from the existing system." The results are then used to "evaluate the conformance of the as-built architecture to the as-documented architecture" to reconstruct
"architecture descriptions for systems that are poorly documented or for
which documentation is not available", and "to analyze and understand the architecture of existing systems to enable modification of the architecture to satisfy new requirements and to eliminate existing software deficiencies."

The reconstruction process derives a representation of software architecture from artifacts such as documentation or, more commonly, the source code or runtime traces. The result assists developers in better understanding the system in question, and it plays a key role in other tasks, such as architecture verification, conformance checking, and trade-off analysis \cite{10.1007/978-3-030-49418-6_21}. Besides these, it is also relevant when facing issues related to software architecture degradation, the process where, due to changes in the codebase, a system's architecture drifts away from the originally intended architecture.

In the context of service-oriented architectures, particularly microservices, the reconstruction process has profound importance and the potential to derive the view of the overall decentralized system. Such a view then shows how the system works \cite{10.1007/978-3-030-49418-6_21}. Obviously, there is a significant difference between assessing monolith systems and decentralized systems like cloud-native microservices. Specifically, the codebase is specific to each cloud-native microservice \cite{TheTwelv82:online}. Each codebase is self-contained and possibly managed possibly by different teams. Moreover, each microservice can follow different conventions, use different versions of libraries, and even follow different platforms. This all makes the reconstruction process more challenging for microservices.

One categorization of reconstruction methods is based on how the analysis is performed. There are three broad groupings: {\em dynamic} or runtime analysis, where a tool constructs the view at runtime; {\em static} analysis, where the view is constructed from artifacts available before deployment~\cite{Cerny2022}; and {\em manual} analysis, where a human examines the system and manually constructs a representation of it~\cite{al2022using}. Manual analysis, while not involving an automated tool, is an important step in establishing a proposed method or validating the results. 

In order to shed light on the existing approaches and technologies for the Software Architectural Reconstruction (SAR) of microservices, in this paper, we provide a review of the different methods available. 

The result of this work could help practitioners and researchers to understand which technique is most suitable for reconstructing microservices-based systems and which tool they can adopt to visualize them. Moreover, the results can also be useful to tool providers that could address the gaps by providing better support for architectural reconstruction or microservices.

This paper is organized as follows. Section II details the dynamic analysis approach, followed by the static analysis approach in Section III. One major outcome of SAR is architectural visualization through views, which we detail and discuss in Section IV, along with current tools for this. We discuss the approaches in Section V and conclude this paper in Section VI.

\section{Dynamic SAR}

Dynamic analysis can operate on several different runtime data sources. It has been used for a myriad of end goals, ranging from analyzing runtime traces extracted from logs to find timing errors \cite{log2018} to extracting service dependency graphs by extracting remote procedure calls from network logs in a microservice mesh \cite{Esparrachiari:2018:TCM:3277539.3277541} or by uncovering dependencies between monitored metrics for components of a distributed system \cite{Thalheim:2017:SAI:3135974.3135977}.  

Runtime analysis-based SAR has taken many forms. One technique is to use instrumentation to insert logging statements 
to report events. It is also possible to use a specialized framework or custom-made annotations in a program, with calls intercepted producing events to be analyzed at runtime. Such events can be used to describe dependencies in microservice systems by detecting where microservices call each other. This approach has been used to model microservice dependencies and find incomplete test coverage of calls across an entire microservice mesh \cite{service_dependency}. This approach can also be used to detect discrepancies between required and provided service versions, as well as generate performance metrics from service error data \cite{version_based_analysis}.

Interceptors can be used to a similar end; Mayer and Weinreich use the Spring framework's interceptors to monitor runtime calls between services to generate an architectural view of a microservice system \cite{interceptor_extraction}. Similarly, calls could be intercepted and rerouted through a security gateway \cite{10.1145/3147213.3147229}, but this can bring significant performance overhead and violation of the distribution with potential bottlenecks. In general, approaches dependent on code instrumentation \cite{interceptor_extraction} bring additional development difficulty and overhead.

Another runtime approach is to build on what we have mentioned above and utilize the underlying containerization engine. Since microservices are often deployed using containers, container configuration files can be a valuable source of information about the application's architecture. Granatelli et al. \cite{recovering_architecture} query the containerization framework to retrieve calls between microservices at runtime. The extracted calls are used along with deployment metadata collected from service descriptors to create an architectural model for a microservice system. This approach is limited in extracting further system concerns because not all information is available through the containerization engine, especially information relating to how the application represents and operates on data.

Finally, the industry practice is to use existing monitoring, tracing, and metrics tools to capture data about the microservices \cite{carnell2021spring}. Established tools recognize enterprise frameworks and utilize existing mechanisms, such as method call interception, instrumentation, or interaction with API-gateway. The additional step to deal with decentralization is that for cloud-native solutions, each logged statement includes correlation ID and origin location. This allows the analytics tools to connect statements related to the same distributed transaction to improve analytics and reasoning. Among industry tool and technology examples, consider OpenTelemetry, Kiali, Zipkin, or Jaeger \footnote{https://istio.io/latest/docs/tasks/observability/kiali/; https://opentelemetry.io; https://zipkin.io; http://jaegertracing.io}. These tools help determine system architecture with specific perspectives they provide from the dynamic information they collect, typically in the form of dependency graphs, directed acyclic graphs, or a topology view. On top of existing tools, others can build additional perspectives. For instance, Khan \cite{khan_2020} constructed the tool MsViz using Jaeger, Grafana, and Prometheus to capture the architecture of microservice systems. He created a graph showing the architecture and overlaid performance data and other metrics onto the graph.

Dynamic analysis approaches benefit from the ability to access runtime data (such as performance metrics and real-time service calls), but they require the system to be deployed, running and used by users. Researchers point out many challenges with distributed tracing. For instance, Bento et al. \cite{tracing21} indicate problems with complexity, application specificity, the volume of information, and lack of tools to abstract, navigate, filter, and analyze trace data in an automated fashion relying on administrators to do it themselves. 

Moreover, with dynamic analysis, we might identify the system endpoints; however, without a complete interaction trace over the system, we will fail to identify the complete system view, and this might be a critical aspect sometimes overlooked when compared to static analysis. For instance, if we intend to perform SAR or microservice correlations and consistency reasoning before production deployment, we need to integrate SAR into the software development lifecycle. Furthermore, unless robust testing and complete test coverage exist, we might not identify all possible communication paths properly and sufficiently. A broader view and precise detail can be uncovered with access to the codebase, but that involves static analysis.

\section{Static SAR}

Static analysis has on great advantage over dynamic analysis. It can be performed on a system before it is deployed, extracting information from existing artifacts that would otherwise have to be manually analyzed.
The most common association with static analysis might be formal verification, where the program's codebases are analyzed to determine the system's correctness \cite{Chlipala:2013:BSP:2544174.2500592, 10.1007/978-3-540-69611-7_8}. 
However, static analysis has been used in automatically generating test cases for a program, for example, by identifying points for performance analysis instrumentation \cite{Cho:2009:UMI:1566445.1566519} or by extracting and analyzing an abstract syntax tree to identify all execution paths that need to be tested \cite{Tonella:2004:ETC:1013886.1007528}. Developers can also use static analysis to better understand a program at a higher level. For example, UML models can automatically be generated by static analysis for legacy systems to better understand how to maintain or replace them \cite{mdd2012}, and it is integral in identifying code clones \cite{Keivanloo:2014:SEB:2844734.2844763, Keivanloo:2012:JBC:2664398.2664404, RATTAN20131165}. 

Static code analysis has also been applied to microservices~\cite{Cerny2022}. It has been used to analyze monolithic systems and recommend splits for converting to microservices \cite{auto_extract_graphs}. Static analysis cannot determine how instances will be instantiated and interact when deriving a precise service dependency graph \cite{Esparrachiari:2018:TCM:3277539.3277541}. However, we must consider that deployment descriptors are available in the codebase indicating remote services by aliases. Next, remote calls are evident in the source code. 
Then, when a system does not contain an indirection (i.e., enterprise service bus), which is the case of microservices. The service dependency graph can be constructed from the source code. Using Java source files and Docker/Spring configuration files have been demonstrated suitable \cite{Pigazzini2020} to reconstruct the architecture of microservices-based systems to identify cyclic dependencies between microservices. Analyzing code to visualize call-graphs between microservices was also proven feasible in \cite{Rahman2019}.

Source code is not the only artifact available for static analysis. It is important to mention that many core artifacts like maven or docker files are typically included in the codebase and can be used for static analysis. Ibrahim et al. use a project's Dockerfiles to search for known security vulnerabilities of the container images being used, which they overlay on the system topology extracted from Docker Compose files to generate an attack graph showing how a security breach could be propagated through a microservice mesh \cite{attack_graph}.
This allows the creation of a centralized security concern for the system, but since it does not extend to source code, it cannot include security flaws in the programs deployed in the containers, only flaws with the images themselves. Another static source of information is in the API definitions. Mayer and Weinreich used API definitions generated by Swagger as an input to their architecture generation system, but their system is also dependent on runtime data extracted from calls between services \cite{interceptor_extraction}.

Another approach to pre-runtime SAR is to embed a source of information in the microservices as part of their development. For example, Salvadori et al. propose creating semantic microservices that expose information about their resources, allowing them to be automatically composed \cite{semantic_microservices}. In this way, a centralized view of microservice communication is always available. However, this approach depends on using a fundamentally different approach to development, and it cannot be used to analyze existing codebases.

\section{Architectural Views} \label{aw}

Software architectures can be described by architectural views \cite{Bass2012}. 
These views capture certain system qualities or aspects. For instance, they are elaborated by the 4+1 architectural view model \cite{Kruchten2020} involving a logical view, process view, development view, physical view, and scenarios, which do not have a visual format. Furthermore, the 4+1 model can be generalized to the N+1 model \cite{4plus1}. 

The foundation for successful SAR is the ability to reconstruct effective architectural views of a system \cite{walker2021automatic}. Existing SAR work related to microservices by Rademacher et al. \cite{10.1007/978-3-030-49418-6_21} has considered four views as their outcome. In particular, it operated with domain, technology, service, and operation views.

\setlist[itemize]{leftmargin=1.3em}
\begin{itemize}[noitemsep,topsep=0pt]
    \item \textit{Domain view} illustrates the domain concepts of in microservice system. More specifically, it describes the data entities of the system along with datasource connections of those entities.
    
    \item \textit{Technology view} details technologies used for microservice implementation and operation.

    \item \textit{Service view} copes with the service models that specify microservice interfaces and endpoints.

    \item \textit{Operation view} then helps the operation to better understand service deployment in the infrastructure. It details containerization, service discovery, and monitoring.
\end{itemize}

Each of these views considered a specific perspective and related concerns within the system. However, each also relates to other views. As an example, consider the service view overlapping with the domain view to detail which data entities are involved in endpoints. The technology and domain view will then show where the data entities persist.

As suggested by Walker et al. \cite{walker2021automatic}, a key point for the construction of these views is that each view is an aggregation of smaller views, each illustrating a disparate microservice. Each microservice can be seen as operating within its bounded context \cite{BoundedC96:online,evans2004domain,vernon2013implementing} of its microservice concerns, but these can be aggregated into a fully centralized perspective of the system's architecture.

\subsection{Views in the Context of Separation of Duty} \label{awd}

These views can be put in contrast to the cloud-native perspectives suggested by Carnell et al. \cite{carnell2021spring} when building microservices. Their book (chapter 3) highlights that "the foundation for successful microservice development starts with the perspectives of three critical roles. The \textit{architect} that sees the big picture of decomposing an application into individual microservices and understanding their interactions.
The \textit{software developer} who codes and understands the language and development frameworks to deliver a microservice (its functions and use cases). The \textit{DevOps engineer} that determines how the services are deployed and managed throughout production and non-production environments". All three roles are essential to ensure proper microservices development. 

Clearly, the architect knows the big picture, while the developer knows one piece of the puzzle in broad detail; finally, the DevOps engineer does not need to know much about what is encapsulated in each microservice but cares about the deployment process, operations, and monitoring. Architectural views should benefit each of these three roles, while their goals and system knowledge are diametrically different. There are, however, overlaps. As of now, it is difficult for the architect to observe whether the developers did their job properly to assure consistency and dependencies across microservices. Proper architectural views would help the architect to ensure the specification was met by the implementation. The DevOps engineers lack a broader understanding of what are the specific needs of the microservice they were asked to deploy. Thus, access to more details (e.g., persistence) would help them to better optimize custom deployment rather than using a one-size-fits-all approach. Finally, developers sometimes need to know what other knowledge microservices encapsulate but might lack time due to the business pressure to assess the specific and resort to wheel reinvention leading to cloned knowledge, business rules, or functionality, introducing technical debt, that sooner or later leads to inconsistencies due to decentralized system evolution leading to architecture degradation.

Moreover, other perspectives can be involved, such as security audits. It is unrealistic to expect one individual to be an expert in all these roles with their different perspectives. Each of these roles needs to have access to distinct architectural views and to assess the system per microservice and holistically. 
Each microservice is developed by developers who are not architects, not DevOps engineers, and likely not security experts. It is unrealistic to expect one individual to be an expert in all these roles, and while there are probably some, generally, we can expect separate experts. Thus, security experts need to assess the system regarding security mechanisms, and privacy and determine weaknesses and vulnerabilities. Similarly, we can account for performance analysis, considering the system for an entirely different goal. Each of these roles needs to have access to distinct architectural views, assess the system per microservice, and also holistically. 

Thus, in the result, we might again refer back to the N+1 model \cite{4plus1}. These views have been detailed in \cite{5676834}, such as \textit{logical view}, \textit{process view}, \textit{deployment/physical view}, \textit{data view}, \textit{security view}, \textit{implementation view}, \textit{development view}, and \textit{use case view}. Not all views can be easily extracted, such as the use case view. 
We detail the other views considered in the N+1 model:
\begin{itemize}
    \item \textit{Logical view} illustrates the conceptual organization of the software in terms of the most important layers, subsystems, packages, frameworks, classes, and interfaces. Typically uses UML package, class, and interaction diagrams.
    
    \item \textit{Process view} details processes and threads, their responsibilities, collaborations, and the allocation of logical elements to them. Typically uses UML interaction and activity diagrams or possibly Business Process Model Notation (BPMN).
    \item \textit{Deployment/Physical view} shows the physical deployment of processes and components to processing nodes and the physical network configuration between nodes. This overlaps with the Operation view suggested earlier by \cite{10.1007/978-3-030-49418-6_21}. UML deployment diagrams are the natural fit for this view.
    \item \textit{Data view} gives an overview of the data flows, persistent data schema, the schema mapping from objects to persistent data (data source), the mechanism of mapping from objects to a database, and database stored procedures and triggers. This partially overlaps with the domain view by \cite{10.1007/978-3-030-49418-6_21} but is more broad considering data flow.
    
    \item \textit{Security view} is an overview of the security schemes and points within the architecture that security is applied, such as HTTP authentication, database authentication, and so forth. It can take advantage of other views starting with the deployment view through the logical view.
    
     \item \textit{Implementation view} details the implementation, these days mostly the specific components involved in the code, packages, and other resources, including other used libraries, which brings an overlap to the technology view.
    
     \item \textit{Development view} detailing the organization of each codebase to help developers orient themselves in the code.
     
     \item \textit{Use Case view} giving a summary of the most architecturally significant use-cases and their non-functional requirements

\end{itemize}


\subsection{Traditional View Modeling}

When considering enterprise architecture, the main emphasis is on business processes. To model enterprise architecture, available languages include ArchiMate, UML, Business Motivation Model (BMM), and BPMN, among others \cite{9097240}. The major focus of UML is objects or components. ArchiMate has derived several concepts from UML but focuses mostly on services. This makes the ArchiMate more suiting to large systems with less detail than when modeling with UML. 

Frameworks for the enterprise architecture practice include The Open Group Architecture Framework (TOGAF). ArchiMate then uses Architectural Development to extend TOGAF. With TOGAF, architectural modeling considers four levels with different specializations: Business, Application, Data, and Technology, which aligns a similarity with Rademacher et al. \cite{10.1007/978-3-030-49418-6_21}. However, the business architecture levels are not covered by Rademacher et al. since this kind of architecture consists of motivation, organization, and mapping of assets, which, while encoded in the system, rather drive the motivation for the implementation.
In particular, these levels are:
\begin{itemize}
    \item \textit{Business architecture} considering governance, organization, the key business processes, actors, services, and qualities when adapting existing processes.
\item \textit{Applications architecture} detailing the individual system deployment including services, logical and physical components, the interactions between the systems, and their relationships to the business processes. (Sharing similarities with the service and technology views).
\item \textit{Data architecture} describing the structure of an organization's logical and physical data assets and the associated data management resources, including data entities. (Similar to the domain view).
\item \textit{Technology architecture} describes the hardware, software, and network infrastructure supporting the deployment of core systems. (Sharing similarities with the technology and operation views).
\end{itemize}

\subsection{Hierarchical View Modeling}
It is recognized that microservices produce complex systems that might need a different level of abstraction when browsing throughout the system. The hierarchical approach of one of the natural choices. The Context, Containers, Components, and Code or just C4 model is worth mentioning as a practical approach for modeling software architecture \cite{c4}. It is a hierarchical model consisting of four levels of abstraction. The high-level system context can let operators transfer to individual code elements. C4 is a natural fit for microservices. While it does not prescribe a method of analysis, its key feature is important for analysis tools to follow in that it allows varying levels of abstraction for the users to see. Such a hierarchical analysis is useful especially for microservices, as it allows views of the system as a whole and inspection of individual services.

Different tools have been proposed to visualize the service call graph. 

\begin{figure}[ht]
\centering
\includegraphics[width=0.5\textwidth]{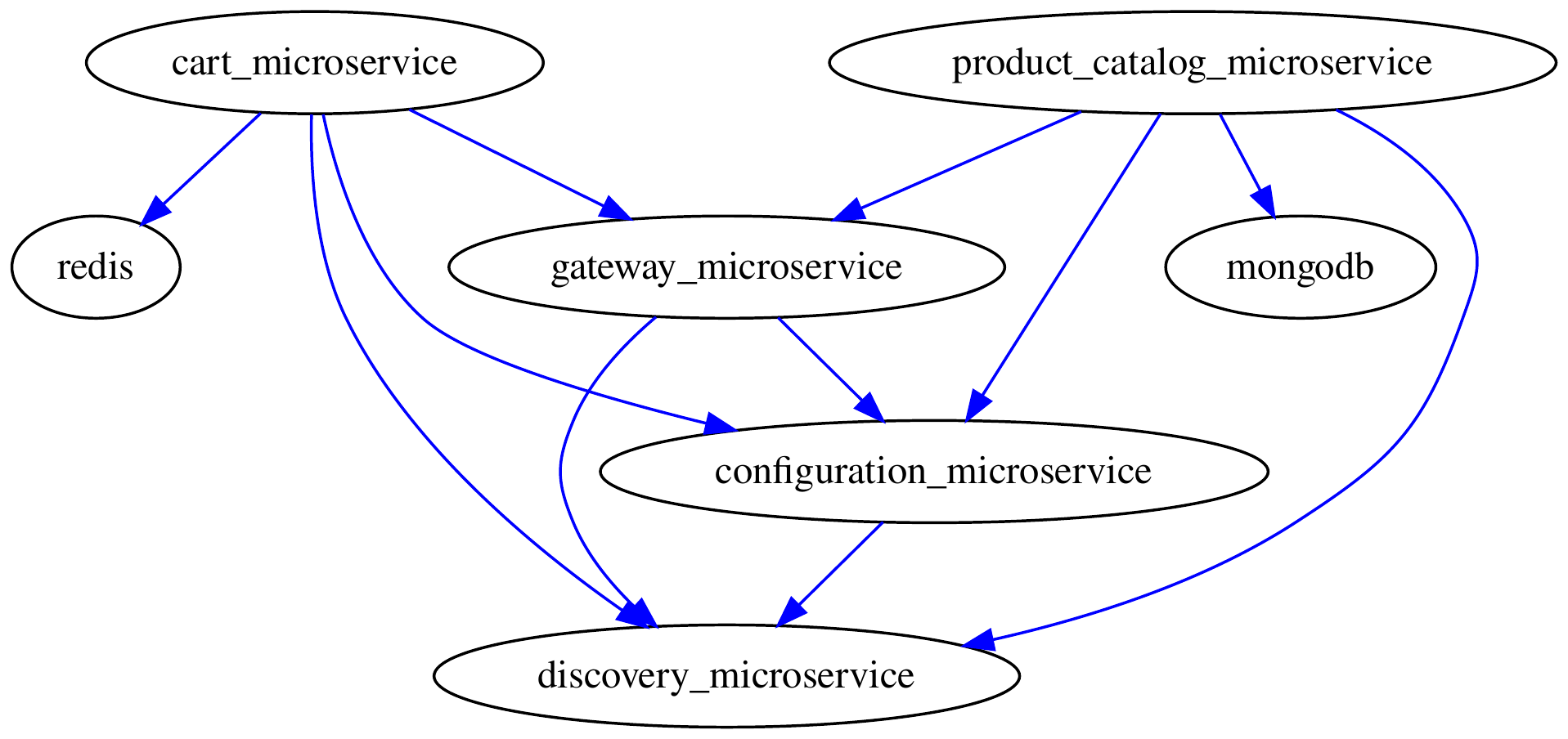}
\caption{MicroDepGraph (from~\cite{Rahman2019})}
\label{fig:microdepgraph}
\end{figure}

Rahman and Taibi~\cite{Rahman2019} proposed a MicroDepGraph\footnote{MicroDepGraph \url{https://github.com/clowee/MicroDepGraph}}, a visualization tool to present the connected services together with the MicroServices Dataset~\cite{Rahman2019}. However, MicroDepGraph does not distinguish between type of services, adopting the same type of shapes to visualize services, databases and other components (e.g. message buses). An example of the visualization proposed by MicroDepGraph is shown in Figure~\ref{fig:microdepgraph}.

\begin{figure}[ht]
\centering
\includegraphics[width=0.5\textwidth]{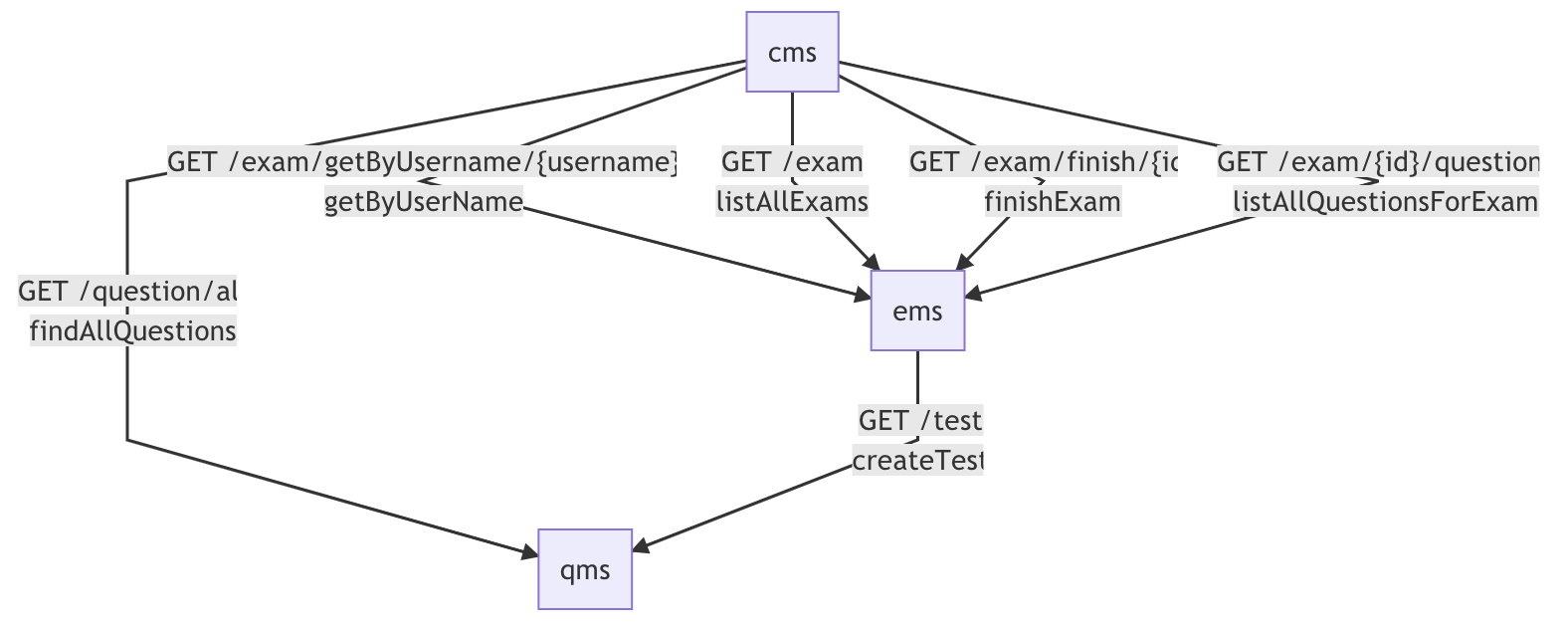}
\caption{Prophet service view (from~\cite{cerny2022microvision})}
\label{fig:prophet}
\end{figure}

In our previous work~\cite{cerny2022microvision}, we developed Prophet\footnote{Prophet:  \url{https://github.com/cloudhubs/prophet-utils},\url{https://github.com/cloudhubs/prophet-utils-app},\url{https://github.com/cloudhubs/prophet}}, a static analysis tool to parse Java-based applications to reconstruct the service call graph of a microservice-based system. Prophet recognizes component-based constructs behind Spring Boot and Enterprise Java. Prophet provides an intermediate graph representation of the system accessible through REST API, to enable system reasoning (Fig~\ref{fig:prophet}).

\subsection{Alternative Visualization for Software Architectures} \label{sec:alternativeVisualiz}

In the area of more general visualization approaches for software architectures \cite{Shahin2014ASR, 10.1145/1409720.1409745} we can operate with various graph-based visualizations showing nodes and edges similar to ontologies, notation-based visualization such as UML or SysML or matrix-based approaches that act as a complementary representation of a graph. Another visualization area of great research interest is metaphor-based visualization. The visualization uses familiar physical world contexts (e.g., cities, islands, or landscapes). 

To select the right visualization category, it has been noted \cite{Shahin2014ASR} that these visualizations often serve a specific purpose. Among the main motivation to visualize architecture is architecture recovery, followed by architecture evolution, impact analysis, general analysis, synthesis, implementation, and reuse.

\begin{figure}[ht]
\centering
\includegraphics[width=0.4\textwidth]{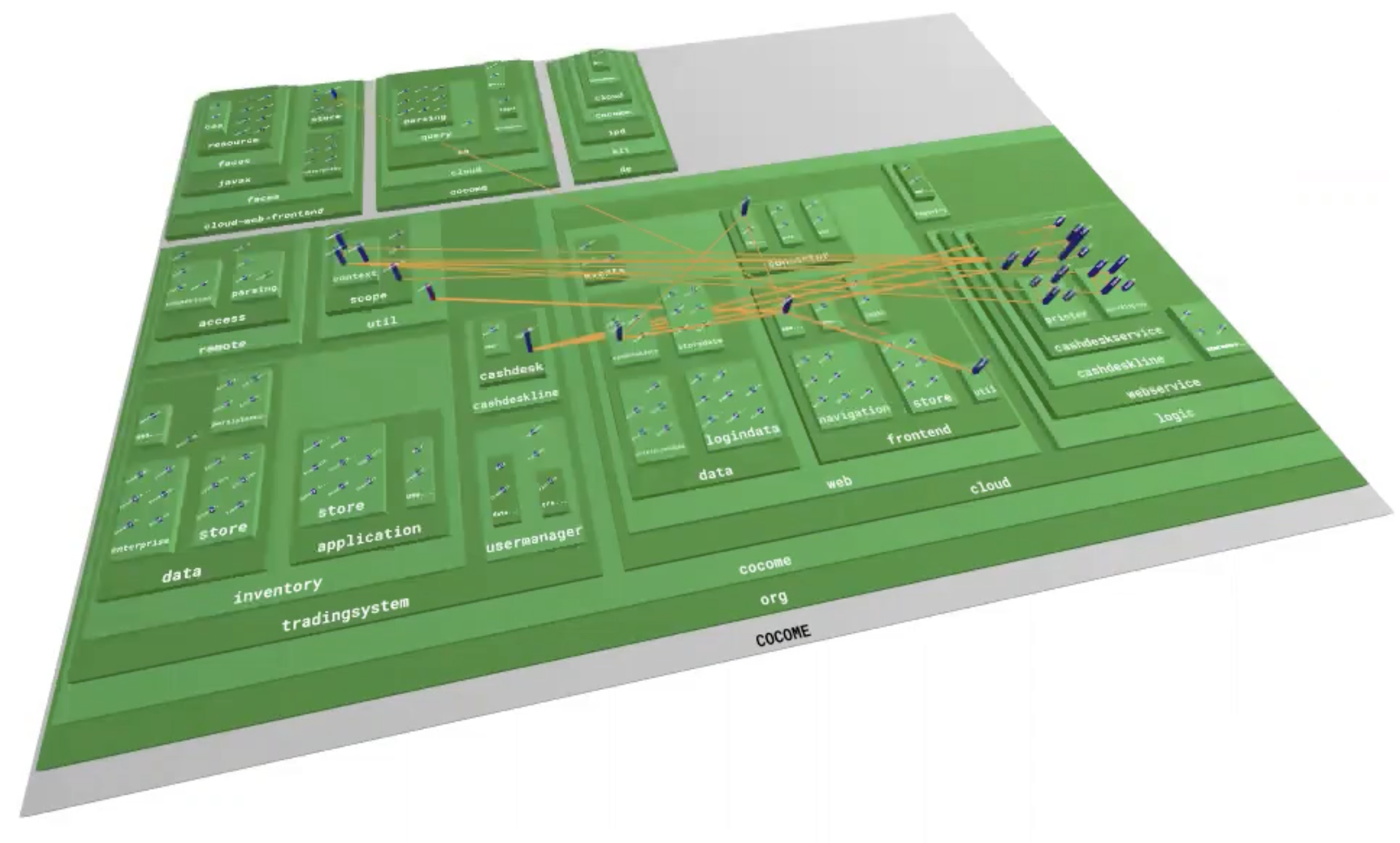}
\caption{The ``software city" metaphor displays an application's package structure as a series of structures building on each other.\cite{vr_software_cities}}
\label{fig:city}
\end{figure}

\begin{figure}[ht]
\centering
\includegraphics[width=0.4\textwidth]{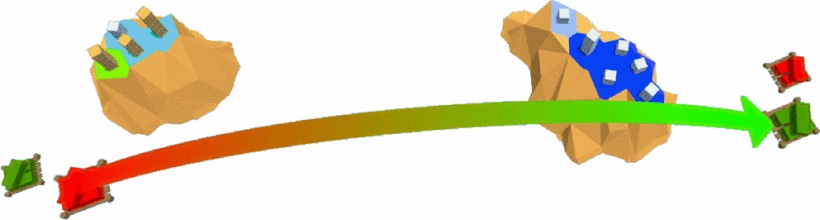}
\caption{Software components can be displayed using the ``software island" metaphor, showing components and their inter-dependencies.\cite{ar_software_islands}}
\label{fig:island}
\end{figure}

The 3D visualization space has been assessed in the literature, but not that much in the context of microservices. In particular, for software architecture visualization, Virtual and Augmented Reality (VR/AR) are good candidates. These can be used as a visual metaphor to make the system more understandable. 
This example is a metaphor for a software city where software packages are represented as buildings and their dependencies as streets. Fittkau et al. implemented the software city metaphor in virtual reality \cite{vr_software_cities} from \fig{city}. Steinbeck et al. \cite{vr_evostreets} presented an even more advanced and scalable derivative called EvoStreets, which gives a better view of the software's hierarchical makeup.
Consider another example shown in \fig{island}. The approach shows individual software modules as islands in an ocean displayed in AR. This visualization method \cite{ar_software_islands} is more closely applicable to a microservice architecture. Software packages and classes in each module are represented as regions and buildings on the module island, and, importantly, module imports and exports are displayed as ports that connect the different islands. While this approach has only been used on monolithic applications, the island metaphor is suitable for displaying the relationships between independent modules in a microservice architecture.

Large microservice-based systems 
are prime candidates for being visualized using VR/AR. One such approach is VR-EA tool \cite{vrea}. Instead of doing dynamic or static analysis to extract a model, VR-EA uses modeling tools as inputs to generate a 3D VR view in the virtual reality of business processes and their relationships with enterprise resources. This approach can provide a comprehensive view of the enterprise system, as it can show a large group of interconnected components. However, it depends on a set of models that must be custom-created to capture the relationships and complexities inside the large system, requiring manual creation of additional configuration and artifacts. 

Virtual reality was used by Ma et al. to monitor a distributed set of servers, visualizing each server as a physical machine in the same VR room \cite{vr_distributed}. Although the monitoring capabilities were limited to system resource usage, the tool showed that physically disparate systems could be virtually collocated to provide a centralized view of a system.

More generally, large systems beyond software architecture have been explored in virtual reality. For example, Toumpalidis et al. used augmented reality to visualize data from IoT networks \cite{vr_binocs}. A user could see a summary of a device's data overlaid on that device. This experiment showed that AR is useful for displaying and aggregating distributed data.

\begin{figure}[t!]
\centering
\includegraphics[width=0.47\textwidth]{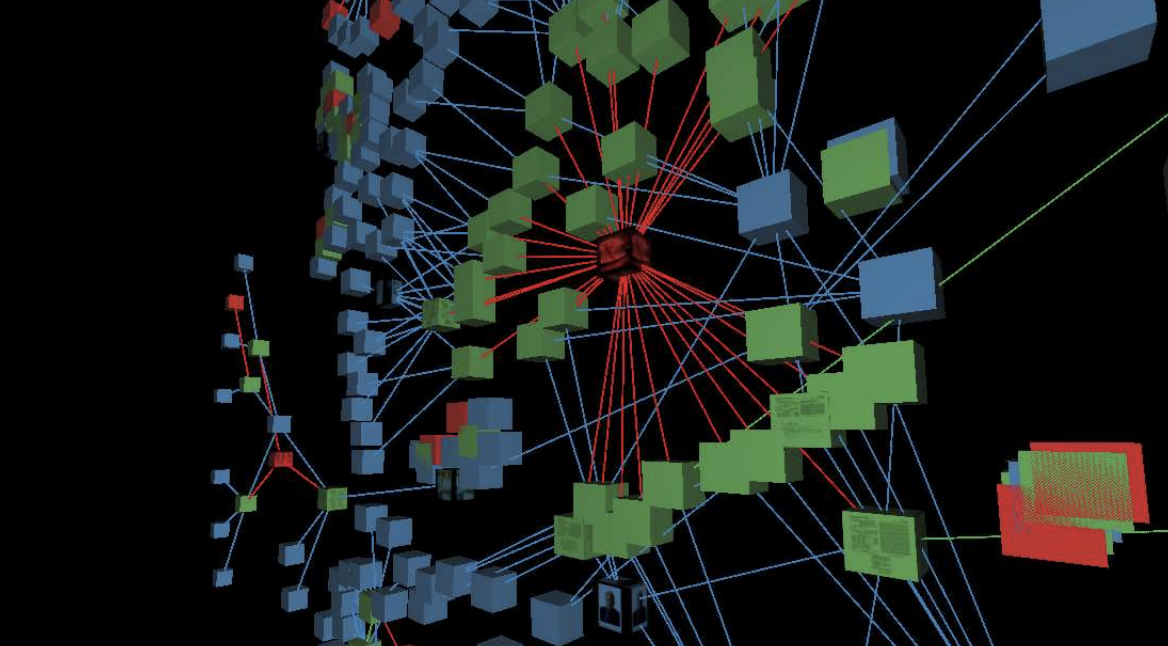}
\caption{Semantic information is a candidate for being displayed in large, three-dimensional graphs due to the natural connections between the elements.\cite{redgraph}}
\label{fig:redgraph}
\end{figure}

Three-dimensional visualizations can also be employed to visualize complex information relationships. Halpin et al. use virtual reality to display the relationships between patent registrants \cite{redgraph} as shown in \fig{redgraph}, and Royston et al. use a similar idea to display connections from social media sites \cite{social}. Neither of these approaches uses a specific visual metaphor, instead opting to display their contents as simple graphs in three-dimensional space. These examples have a common structure with the microservice architectures, as microservices architecture can be viewed as a network of services communicating with each other based on semantic relationships.



Moreover, Moreno-Lumbreras et al. \cite{morenolumbreras2021vr} have suggested using VR to visualize development metrics and analytics in three-dimensional space. Acquiring a broad range of aspects and views for this kind of visualization may be regarded as similar to architectural reconstruction. 
They analyzed and compared the comprehension of metrics in code reviews when aided by VR or 2D visualization. However, no results are available yet.

\begin{figure}[t]
\centering
\includegraphics[width=.47\textwidth]{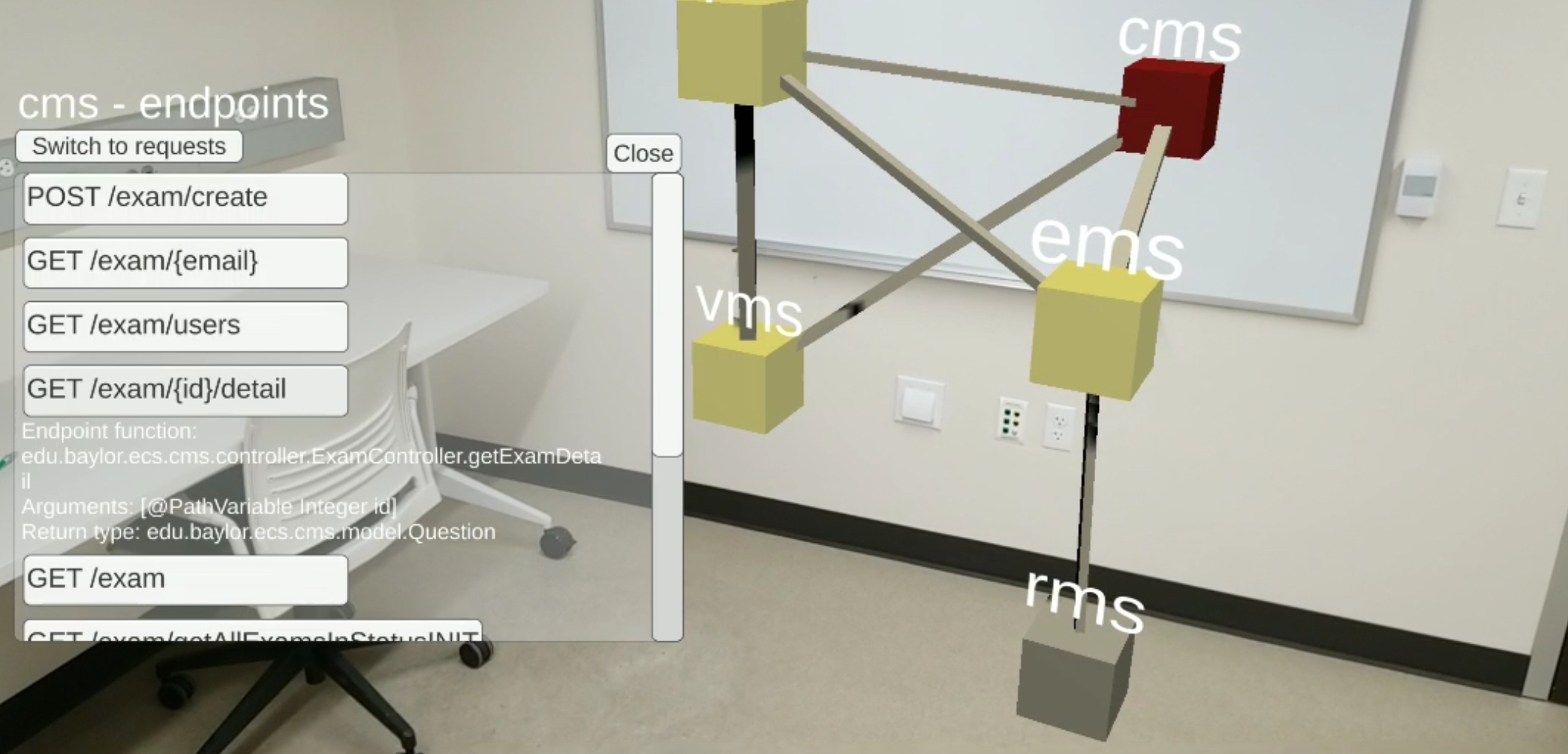}
\caption{Microvision Augmented Reality Visualization. The context menu shows a selected services API endpoints, in this case the ``cms" service highlighted in red.}
\label{fig:microvision}
\vspace{-2em}
\end{figure}

In our previous work~\cite{cerny2022microvision}, we developed and we empirically evaluated Microvision, an AR tool that uses the intermediate representation of the system built by the static analysis tool Prophet, which is capable of multi-codebases analysis for microservices to visualize the service endpoints of microservices, confirming the usefulness of AR for system understanding. An example of Microvision visualization is reported in Figure~\ref{fig:microvision}. 

\subsection{Microservice visualization in the industry} \label{sec:industry}

Microservice development comes from practitioners, and research tends to come later, so publications about microservices are still limited in a lot of areas.
Thus, grey literature may hold valuable insights that academic literature simply cannot provide yet \cite{Bogner2021}.

\begin{figure}[h!]
\includegraphics[width=3.2in]{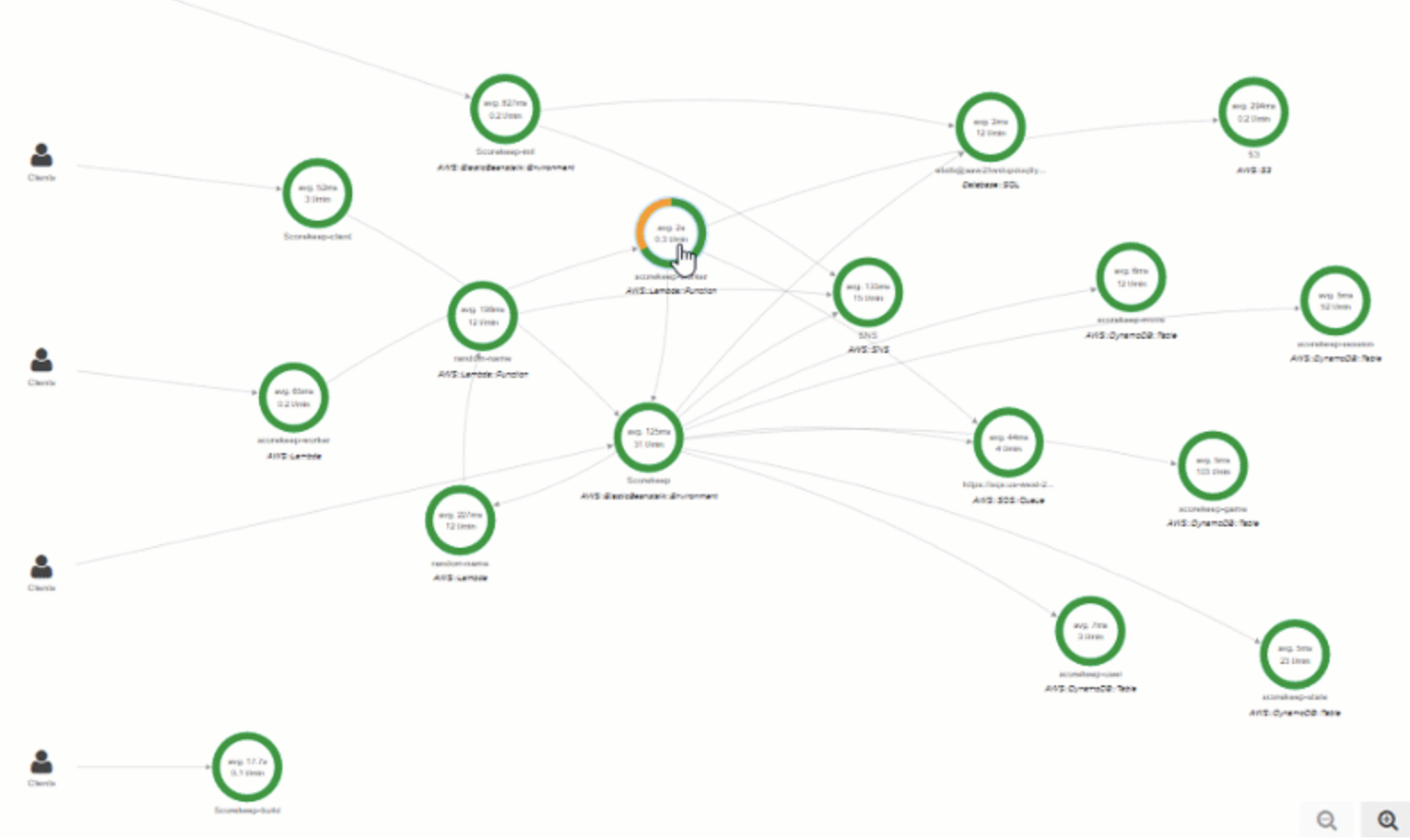}
\caption{Amazon X-Ray Console for Microservices Visualization}
\label{fig:xray}
\end{figure}

Amazon provides a solution called X-Ray console\footnote{\url{https://aws.amazon.com/xray/}}. The provided approach is a map visual representation that consists of service nodes that serve requests, upstream client nodes that represent the origins of the requests, and downstream service nodes that represent web services and resources used by an application while processing a request (as depicted in \fig{xray}). The X-Ray console provides embedded views that enable the user to view service maps and traces of applications' requests.

\begin{figure}[h!]
\includegraphics[width=3in]{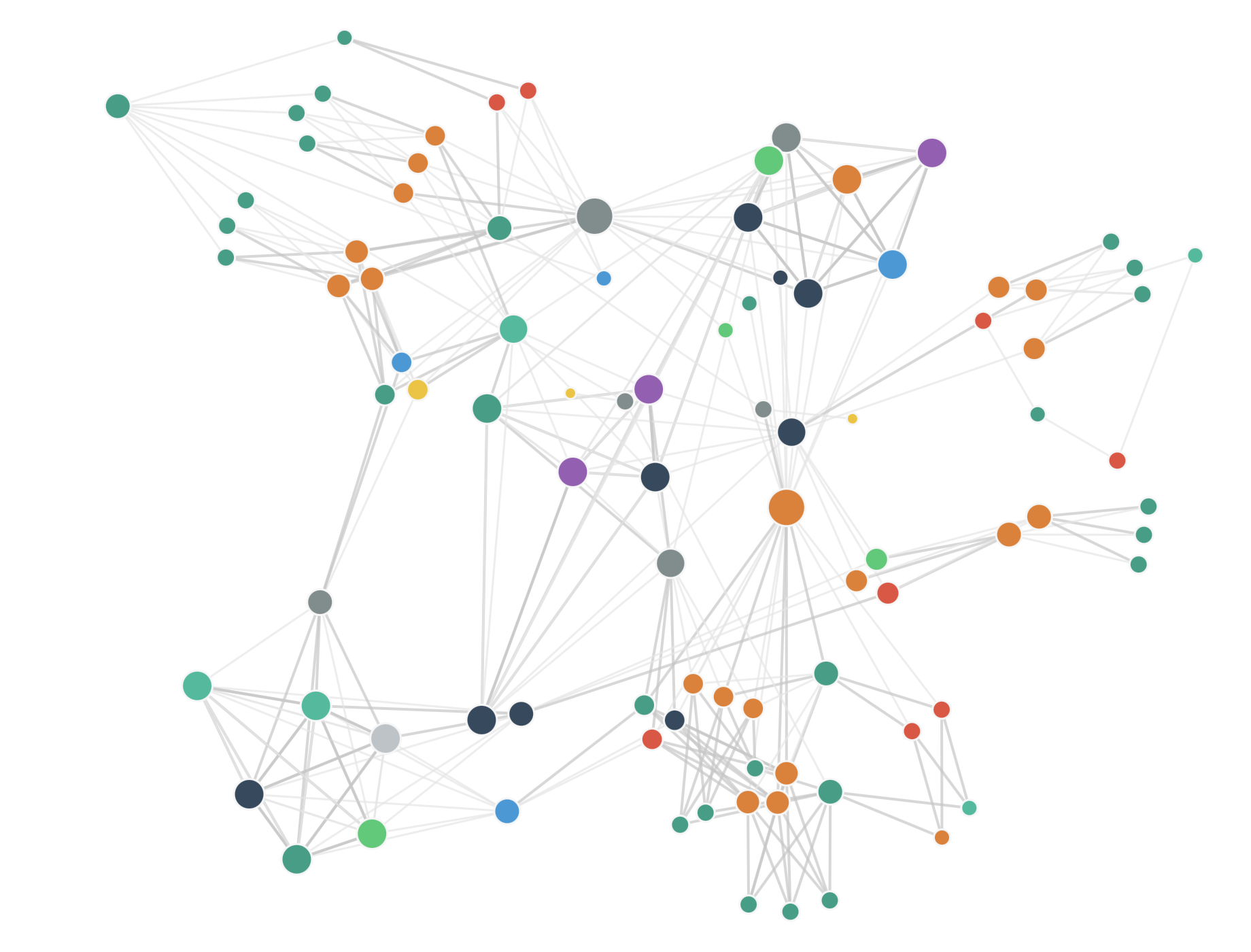}
\caption{Netflix Interactive Microservices Visualization}
\label{fig:netflix}
\end{figure}

Netflix provides an interactive visualization technique for their system\footnote{\url{http://simianviz.surge.sh/netflix}}. 
\fig{netflix} shows the service graph representation of the system. It illustrates service dependencies in the whole system and enables the user to reconstruct the services communication graph to analyze different topologies. However, such a topology view is not particularly useful in debugging where a specific service is experiencing an issue.

\begin{figure}[h!]
\centering
\includegraphics[width=1.5in]{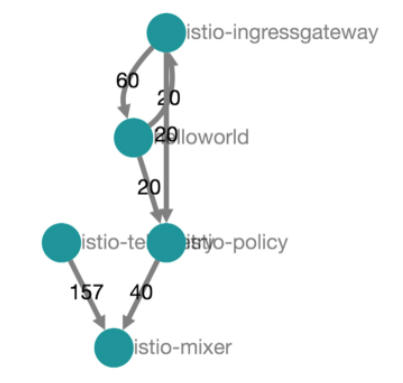}
\caption{Jaeger's directed acyclic graph with call frequencies}
\label{fig:jaeger}
\end{figure}

Jaeger tracing\footnote{\url{https://www.jaegertracing.io}} is a common tool using dynamic analysis that provides Jaeger UI to render service dependencies. \fig{jaeger} shows a visual Directed Acyclic Graph (DAG) from Jaeger UI along with frequencies of calls. It can render a view to observe the system architecture.

\begin{figure}[h!]
\includegraphics[width=3.5in]{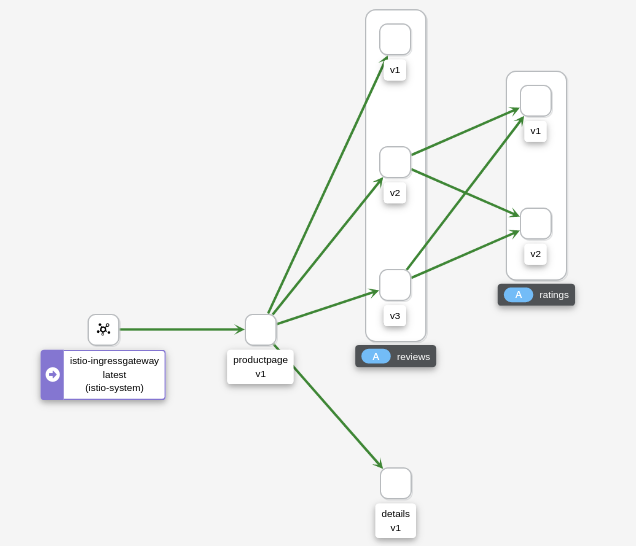}
\caption{Kiali traffic flowing through the service mesh}
\label{fig:kiali}
\end{figure}

When using the istio service mesh, a visualization is provided by the Kiali mesh visualization tool \footnote{\url{https://istio.io/latest/docs/tasks/observability/kiali/}}. Kiali produces graphs representing traffic flowing through the service mesh for a period of time. There are several graph types provided, such as application, versioned application, workload, or service. Application type aggregates all versions of an app into a single graph node. The Versioned application graph type shown at \fig{kiali} shows a node for each version of an app, but all versions of a particular app are grouped together.
The workload graph type shows a node for each workload in the service mesh. Finally, the service graph type shows a high-level aggregation of service traffic in the mesh.


\section{Discussion}
\label{sec:discussion}

Different approaches have been proposed for SAR. 

As for SAR-based on dynamic analysis, the industry seems to be a step forward compared to research. Industry proposed different visualization tools, mainly based on the service call-graph. However, no visualization provides insight into the quality of the architecture or its possible degradation. 

Dynamic analysis might be very useful for making business decisions on the system and the development priority. As an example, companies might prioritize the maintenance of service more used by their customers; it would be possible to compare the maintenance effort with the actual usage of a service, but also to understand which service can be removed from the system. 
However, existing tools do not currently provide support for these decisions.  However, An important lack of dynamic analysis tools is the support for architectural patterns~\cite{Taibi2018closer18}, anti-patterns~\cite{Taibi2020MSE}, and software metrics (e.g. Coupling~\cite{Panichella2021}). For example, the service call-graph enables the detection of different patterns~\cite{bakhtin2022survey},  anti-patterns and to calculate metrics such as coupling and cohesion. Therefore, we recommend tool providers introduce such features to provide better support to companies using their tools.

SAR based on static analysis received more attention from researchers, with little consideration from the practitioner's point of view. However, we would like to stress the importance of architectural reconstruction with static analysis. The main reason is that issues detected with static analysis could be immediately notified to the developers before the system goes into production. As an example, it could be possible to set quality gates in pull requests in case the developers introduce some architectural anti-pattern.


\section{Conclusions} \label{sec:conclusion}

This research was motivated by recurrent microservice system challenges regarding missing system-centric views. In addition, the microservices decentralization nature leads to misaligned documentation and, consequently, architectural degradation, inefficiency, and broadening efforts.

Current approaches to determine the system-centric perspective prioritize dynamic analysis, targeting technology-neutrality. While it helps DevOps with their task, it does not necessarily fit developers as it requires dynamic system interaction. While user simulation tests can be developed, these easily degrade as the system evolves. 

This work investigates the existing methods for static and dynamic architectural reconstruction and the tools adopted to visualize it. 

Results show that static analysis is still not properly developed and might deserve special attention from the practitioner's point of view. Dynamic analysis, instead, is now widely used by several tools to visualize the service call-graph. However, there is no support for architectural degradation or investigation of quality issues in microservices. 

Future work includes the development of extensions for dynamic analysis tools to detect architectural smells and to calculate software metrics such as coupling and cohesion. Moreover, we are planning to survey developers to understand their needs in terms of architectural degradation and further develop static and dynamic analysis tools based on developers' feedback.

\section*{Acknowledgments}
This material is based upon work supported by a National Science Foundation under Grant No. 1854049, a grant from Red Hat Research (https://research.redhat.com), and support from the Shapit Project (Ulla Tuominen Foundation - Finland).

\bibliographystyle{IEEEtran}
\bibliography{access}

\end{document}